\documentclass[12pt,a4paper,oneside]{article}
\usepackage{graphicx,amssymb,amsmath,color,xcolor}
\usepackage[linkcolor={blue},citecolor={red},colorlinks=true]{hyperref}
\usepackage{enumerate}
\usepackage[margin=2 cm]{geometry}
\usepackage{slashed}
\usepackage{multirow}
\usepackage[normalem]{ulem}

\def\beq{\begin{equation}}
\def\eeq{\end{equation}}
\def\bea{\begin{eqnarray}}
\def\eea{\end{eqnarray}}

\def\bit{\begin{itemize}}
\def\eit{\end{itemize}}

\def\l{\left}
\def\r{\right}

\def\baa{\begin{array}}
\def\eaa{\end{array}}

\def\sl#1{\mathord{\not\mathrel{{\mathrel{#1}}}}}
\def\d{\partial}

\def\simgt{\mathrel{\lower2.5pt\vbox{\lineskip=0pt\baselineskip=0pt
           \hbox{$>$}\hbox{$\sim$}}}}
\def\simlt{\mathrel{\lower2.5pt\vbox{\lineskip=0pt\baselineskip=0pt
           \hbox{$<$}\hbox{$\sim$}}}}
\newcommand{\vev}[1]{ \langle {#1} \rangle }

\def\bfc{\begin{figure}\begin{center}}
\def\efc{\end{center}\end{figure}}
\def\nn{\nonumber\\}

\newcommand{\red}[1]{{\color{red} #1}}

\begin{document}

\begin{flushright}
\hspace{3cm} 
SISSA 13/2021/FISI\\
TU-1127
\end{flushright}
\vspace{.6cm}
\begin{center}

\hspace{-0.4cm}{\Large \bf 
Baryogenesis via  relativistic bubble walls}\\[0.5cm]

\vspace{1cm}{Aleksandr Azatov$^{a,b,c,1}$, Miguel Vanvlasselaer$^{a,b,c,2}$ and Wen Yin$^{d,e,3}$}
\\[7mm]
 {\it \small

$^a$ SISSA International School for Advanced Studies, Via Bonomea 265, 34136, Trieste, Italy\\[0.15cm]
$^b$ INFN - Sezione di Trieste, Via Bonomea 265, 34136, Trieste, Italy\\[0.1cm]
$^c$ IFPU, Institute for Fundamental Physics of the Universe, Via Beirut 2, 34014 Trieste, Italy\\[0.1cm]
$^d$ Department of Physics, Tohoku University,  
Sendai, Miyagi 980-8578, Japan \\[0.1cm]
$^e$ Department of Physics, The University of Tokyo,   
Bunkyo-ku, Tokyo 113-0033, Japan\\[0.1cm]
 }
\end{center}

\bigskip \bigskip \bigskip

\centerline{\bf Abstract} 
\begin{quote}
We present a novel mechanism which leads to the baryon asymmetry generation during the strong first order phase transition. If the bubble wall propagates with ultra-relativistic velocity, it has been shown \cite{Vanvlasselaer:2020niz} that it can produce states much heavier than the scale of the transition and that those states are then out of equilibrium. 
In this paper, we show that this production mechanism can also induce CP-violation at one-loop level. We calculate those CP violating effects during the heavy particle production and show, that combined with baryon number violating interactions, those can lead to successful  baryogenesis. Two models based on this mechanism are constructed and their phenomenology is discussed. Stochastic gravitational wave signals turn out to be generic signatures of this type of models.

\end{quote}

\vfill
\noindent\line(1,0){188}
{\scriptsize{ \\ E-mail:
\texttt{$^1$\href{mailto:aleksandr.azatov@NOSPAMsissa.it}{aleksandr.azatov@sissa.it}},
\texttt{$^2$\href{miguel.vanvlasselaer@NOSPAMsissa.it}{miguel.vanvlasselaer@sissa.it}},
\texttt{$^3$\href{yin.wen.b3@tohoku.ac.jp}{yin.wen.b3@tohoku.ac.jp}}
}}

\newpage

\newpage
\section{Introduction}

 One of the greatest puzzles of the early universe cosmology is  \red{the} origin of  the observed excess of matter over anti-matter. The over-abundance of matter is commonly parametrized via the \emph{baryon-to-photon} density ratio 
 \bea
 Y_{ B} \equiv \frac{n_B -n_{\bar{B}}}{s}\bigg|_0
 \eea
 with $n_B, n_{\bar{B}}$ and $s$ respectively the number density of baryons, anti-baryons and the entropy density. The subscript 0 means ``at present time". Planck data and evolution models of the early universe permit to compute this ratio with 
 high accuracy\cite{Ade:2015xua}
\bea
Y_B = (8.75 \pm 0.23)\times 10^{-11}.
\eea  
 Though this ratio is much smaller than unity, it calls for an explanation in terms of early universe dynamics, i.e. baryogenesis. The necessity  of baryogenesis aggravates in the inflationary cosmology since this asymmetry ratio cannot be attributed to the initial conditions.

For a successful baryogenesis scenario,  the 
well-known Sakharov requirements should be 
satisfied~\cite{Sakharov:1967dj}
namely the 
violation of the baryon number, violation of 
$C$ and $CP$ symmetries, and the presence of an out-of-equilibrium process. Based on this
general requirements various models have been constructed (for reviews see for example
\cite{Riotto:1998bt, Bodeker:2020ghk}) based on the different realizations of the Sakharov's conditions.
One interesting possibility for the fulfillment of the out-of-equilibrium process requirement is a scenario in which a first order phase transition (FOPT) occurs in the early history of the universe and this will be the focus of the study in the present paper.
However 
in the standard model the phase transitions are not of the first order (neither QCD PT\cite{Bhattacharya:2014a} nor electroweak one \cite{Kajantie:1996mn}).
This fact prevents the realization, within the SM, of the very attractive idea of \emph{electroweak baryogenesis}
 \cite{Kuzmin:1985mm, Shaposhnikov:1986jp}.  This implies that  physics beyond the standard model (BSM) is needed.

Various physically motivated extensions  of the standard model,~like MSSM or composite Higgs models, could provide room for baryogenesis during the electroweak phase transition (EWPT) \cite{Nelson:1991ab,Carena:1996wj,Cline:2017jvp,Bruggisser:2018mrt,Bruggisser:2018mus} (see \cite{Morrissey:2012db} for review) or some other phase transitions \cite{Long:2017rdo} in the early universe. Interestingly, in most of the cases, the successful generation of the baryon asymmetry 
requires the slow motion of the bubble walls (
though bubble velocity can be supersonic \cite{Caprini:2011uz,Cline:2020jre,Dorsch:2021ubz}
or in the case of specific models even relativistic
\cite{Katz:2016adq}.
In this paper we propose a 
new mechanism for the 
production of the baryon 
asymmetry during the first 
order phase transition, which is only effective in the opposite regime i.e. for the ultra-relativistic bubble wall expansions.
The 
idea will be based on the 
recent observation in
\cite{Vanvlasselaer:2020niz}
, where it was
shown that  in the presence 
of a ultra-relativistic bubble expansion, with Lorentz 
factor $\gamma_w \gg 1$, particles with mass 
up to $M \lesssim \sqrt{\gamma_w 
T_{\text{nuc}} \times v}$ can be produced.
The parameters $T_{\text{nuc}}$ and $v$  are 
the temperature of FOPT (nucleation temperature) and the scale of the 
symmetry breaking respectively. The process of 
the heavy states production during the FOPT is 
obviously out-of-equilibrium, so that if it 
proceeds through a CP-violating and process and 
baryon number is not preserved we can have a 
successful baryogenesis scenarios\footnote{For 
other baryogenesis models with new heavy fields 
production during FOPT see \cite{Katz:2016adq}.}. %
We confirm the statements above by analyzing the  CP-violating effects in the interference of tree and one loop level processes. Then we construct
explicit models where the baryogenesis is realized during the strong FOPT, 
which can be either the EWPT itself, 
if it comes with the necessary new 
physics, or related to some other 
symmetry breaking in the early universe.

One of the interesting feature of  this class of models is that, contrary to  the ``traditional" baryogenesis models, it needs ultra-relativistic bubble wall velocities and 
is generically accompanied with 
strong gravitational waves signal.

The remaining of the paper is organized as follows: in the section \ref{sec:CPviol}, we review the mechanism of the heavy state production proposed in \cite{Vanvlasselaer:2020niz}
and calculate the CP violation in this process. In the section \ref{sec:applications} we build two models 
of baryogenesis and discuss their phenomenology and then in the section \ref{sec:summary} we conclude by recapitulating the main results of this work.

\section{Mechanism of CP-violation via bubble wall}
\label{sec:CPviol}
\subsection{Production of the heavy states in the phase transition}
Let us start by reviewing the process of heavy states production during the phase transition presented in \cite{Vanvlasselaer:2020niz}. We will assume that the phase transition is of first order and that the bubbles reach ultra-relativistic velocities during the expansion $\gamma\gg 1$. We discuss the conditions for such a dynamics in Appendix \ref{app:dynamics}. To make the discussion  explicit we assume the following Lagrangian:
\bea 
\mathcal{L} =
|\d_\mu \phi|^2+i \bar \chi \sl\d \chi + i \bar N \sl\d N- M \bar N N- Y \phi \bar N \chi
\label{eq:Toy_model_1}-V(\phi)
\eea
where  $\phi$ is a scalar field
undergoing a FOPT of e.g. global 
$U(1)$ symmetry, $\chi$ a light
fermion and $N$ a heavy Dirac fermion with mass $M\gg \vev{\phi}$, $M\gg T_{\text{nuc}}$ and $Y$ is the coupling between the scalar and the two fermions. $V(\phi)$ is a potential for the field $\phi$, which we will assume leads to the FOPT without specification of its explicit form. Here and hereafter, without loss of generality,  we work in the basis where fermion masses are real. So that before, during and after the FOPT the equilibrium abundance of $N$ is exponentially suppressed. However in the case of an ultra-relativistic bubble expansion, the probability that the light $\chi$ fluctuates via mixing to the heavy $N$ is non-vanishing \cite{Vanvlasselaer:2020niz} and is approximately equal to
 \bea
 \mathcal{P}^{tree}(\chi \to N) \approx  \frac{Y_{}^2 \vev{\phi}^2}{M^2}\Theta(\gamma_w T_\text{nuc} - M^2 L_w)
 \label{eq:prod_tree}
 \eea
with $L_w \sim 1/\vev{\phi}$ the
length of the wall.
Thus, when the ultra-relativistic wall 
hits the plasma, it produces $N$ and $\bar{N}$. Note that this abundance will be much larger than its equilibrium value. 

\subsubsection{Method of calculation of the light $\to$ heavy transition}
Before we proceed to the one loop calculation, let us present a generic method to compute the transition amplitudes\footnote{We are using slightly different derivation compared to the original paper \cite{Vanvlasselaer:2020niz}}. We will then apply it to recover Eq.\eqref{eq:prod_tree} and later for the computations of the one loop corrections. In this section we will omit the flavour indices which we will easily recover once the loop functions will be derived.

    Let us look at the correlation function
$\langle 0| T\{ \bar \chi(x_1)N(x_2) \}|0\rangle$ 
and calculate it to first order in ${\cal O}\l(\frac{\vev{\phi}}{M }\r)$ which will be our expansion parameter. We assume that the wall is located in $x-y$ plane at $z=0$. The correlation functions writes 
\bea
\langle 0| T\{ \bar \chi(x_1)N(x_2) \}|0\rangle=
\int d^4 x Y \vev{\phi(x)} S_\chi(x_1-x) S_N(x-x_2) + \mathcal{O}\l(\frac{Y\vev{\phi}}{M}\r)^2
\eea
where we are expanding the correlation functions of the theory with $\vev{\phi}\neq 0$ in terms of the correlation functions $S_{\chi,N}$ of the unbroken $\vev{\phi}=0$ theory. Then performing the Fourier transformation we will obtain
\bea
&&\int d^4 x d^4 k d^4q e^{ik(x_1-x)+iq(x-x_2)}S_\chi(k)S_N(q)Y\vev{\phi(x)}\nn
&&=\int d^4 k d^4q e^{i k x_1- i q x_2} S_\chi(k)S_N(q)  \times \l[(2\pi)^3 \delta^{(3)}(k-q) \int d z e^{i z(k_z-q_z)}
Y\vev{\phi(z)}\r],
\nn
&&\delta^{(3)}(k-q)\equiv \delta^{(1)}(k_0-q_0)\delta^{(1)}(k_x-q_x)\delta^{(1)}(k_y-q_y) .
\eea
Let us make a  few comments regarding this expressions. The propagators $S_{\chi,N}$ have poles at, respectively $p^2=0, M^2$ and this, together with energy and $(x-y)$-momentum conservation, fixes the exchange of momentum $\Delta p_z$ from the plasma to the wall;
\bea
\Delta p_z = q_z-k_z=-k_z+\sqrt{k_z^2-M^2} \approx -\frac{M^2}{2k_z}.
\eea
 Now we can use the LSZ reduction formula to relate the correlation function to the matrix element  of the $\chi\to N$ transition and we find that 
\bea
\langle N,q|\chi,k \rangle= \l[(2\pi)^3 \delta^{(3)}(k-q)\int d z e^{-i z\Delta p_z}
\vev{\phi(z)}\r]\times  \bar u_N(q)u_\chi(k) Y
\eea
which coincides with the result found in\cite{Vanvlasselaer:2020niz}. Note that the last factor is exactly equal to the amplitude of the transition $\chi(k)\to N(q) \phi(\Delta 
p_z)$, ${\cal M}_{\chi(k)\to N(q) \phi(\Delta 
p_z)}$.
 Thus we can write
\bea
\label{eq:mastertrans}
\langle N,q|\chi,k \rangle= \l[(2\pi)^3 \delta^{(3)}(k-q)\int d z e^{-i z\Delta p_z}
\vev{\phi(z)}\r] {\cal M}_{\chi(k)\to N(q) \phi(\Delta 
p_z)}.
\eea
Of course on-shell $\phi$ cannot have the space-like momentum $\Delta p_z$ but since it is a scalar we can still formally define such an ``amplitude". These relations are the consequence of the following \emph{Ward identity}  which is satisfied if we are looking at the effects with just one VEV $\vev{\phi}$ insertion:
\bea
\langle O_1(x_1)...O_n(x_n) \phi(x_{n+1})\rangle|_{\vev{\phi}=0} =\int d z \l  [ \frac{\delta }{\delta \vev{\phi(z)}} 
\langle O_1(x_1)...O_n(x_n)\rangle| _{\vev{\phi}\neq 0}\r]D_\phi(x_{n+1}-z),
\eea
where $D_\phi$ is the propagator of the $\phi$ field.
Then the application of the LSZ reduction together with energy and transverse momentum conservation  leads to the Eq.\eqref{eq:mastertrans}.

\subsubsection{Probability of the light $\to$ heavy transition}
Armed with the generic expression in Eq. \eqref{eq:mastertrans} we can proceed to the computation of the light $\to $ heavy transition probability similarly to the discussion in  \cite{Vanvlasselaer:2020niz,Azatov:2021ifm},
\bea
P_{\chi\to N}=\int \frac{d^3 q}{(2\pi)^3 2 q_0 2 k_0}(2\pi)^3 \delta^{(3)}(k-q)|{\cal M}_{\chi(k)\to N(q) \phi(\Delta 
p_z)}|^2\l|\int dz e^{-i z \Delta p_z}\vev{\phi(z)}\r|^2.
\eea
Performing the phase space integral and summing (averaging) over incoming (outgoing) spins, we arrive at
\bea
P_{\chi\to N}=\l|\int dz e^{-i z \Delta p_z}\vev{\phi(z)}\r|^2\times \frac{|Y|^2k_z (k_z-\sqrt{k_z^2-M^2})}{2k_0\sqrt{k_z^2-M^2}}
\label{eq:transition-exact}
\eea
Let us evaluate the first prefactor which takes into account the shape of the wall. To approximate the integral, we need to use some estimation for the shape of the wall. For example for a \emph{linear} wall ansatz  of the form
\bea
 \langle \phi(z)\rangle =\l\{\baa{c} 0,~~z <0\\
 \langle \phi\rangle\frac{z}{L_w}  ~~~0\leq z\leq L_w\\
 \langle \phi\rangle ~~~z>L_w
 \eaa\r .
  \eea
(where we use $\langle \phi\rangle$ as the VEV of $\phi$ in the true vacuum) we obtain
   \bea \l|\int d z e^{-i z\Delta p_z}
\vev{\phi(z)}\r|^2& =& \frac{\vev{\phi}^2}{\Delta p_z^2} \bigg(\frac{\sin \alpha}{\alpha} \bigg)^2 , \qquad \alpha = \frac{L_w \Delta p_z}{2}.
 \label{eq:matrix_ele}
  \eea
  We thus observe the appearance of a new suppression factor that becomes relevant in the limit $\alpha =\frac{L_w\Delta p_z}{2} \gg 1$ and quickly suppresses the transition. Similar results holds for more realistic wall shapes when it is given by either $\tanh$ or gaussian functions
\bea
 \langle \phi\rangle_{\text{tanh}}(z)= \frac{ \vev{\phi}}{2} \bigg[\tanh{\l(\frac{z}{L_w}\r)} +1\bigg],
\qquad 
 \langle \phi\rangle_{\text{gaussian}}(z)= \frac{ \vev{\phi}}{\sqrt{2 \pi} L_w } \int_{-\infty}^{z}d z'\exp{\l(-\frac{z'^2}{2L_w^2}\r)}.
\eea
In these cases we find respectively (see for details of calculation \cite{Azatov:2021ifm}):
\bea
\l|\int d z e^{-i z\Delta p_z}
\vev{\phi(z)}\r|^2_{\text{tanh}}&=&\l[\frac{ 
\pi    L_w }{2 \sinh{\l( \frac{ L_w \Delta p_z \pi}{2}\r) }}\r]^2\vev{\phi}^2,\nn
\l|\int d z e^{-i z\Delta p_z}
\vev{\phi(z)}\r|^2_{\text{gaussian}}&=& \frac{ \vev{\phi}^2 }{ \Delta p_z^2} \exp{\l( -L_w^2\Delta p_z^2 \r)}.
\eea
Thus we can see that independently of the wall shape  the transitions with $\Delta p_z \gg L_w^{-1} $ are strongly suppressed. In the opposite regime $\Delta p_z \lesssim L_w^{-1}\sim \vev{\phi}$ we will have $k_z\gtrsim M^2/\vev{\phi}$. Then expanding the Eq. \eqref{eq:transition-exact}, we will obtain
\bea
P_{\chi\to N}\simeq \frac{Y^2 \vev{\phi}^2}{M^2}\Theta( k_0 - M^2 L_w),
\label{eq;trans_light_heavy}
\eea
which reduces to Eq.\eqref{eq:prod_tree} when we notice that $k_0 \sim \gamma_w T_{\rm nuc}$.
We can see that indeed there will be an efficient production of the heavy states, which will not be Boltzmann suppressed, however the Lorentz boost factor $\gamma_w$ for the wall expansion needs to  be large enough.

After this warm-up exercise  we can proceed to the calculation of the one loop effects. 
We will focus again only on the terms with just one VEV $\vev{\phi}$ insertion and proceed in the same way as we
have done for the tree level calculation. Note that the Eq. \eqref{eq:mastertrans} will remain true also at loop level if we are focusing only on the effects with one VEV
 insertion. Indeed the momentum is not conserved only in the vertex with the $\vev{\phi}$ insertion, however the energy and $x-y$ momentum conservation still fixes the value of the loss of the $z$ component of momentum. At this point since $\phi$ is a scalar (no polarization vectors are needed) the matrix element is exactly the same as for the process  $\chi(k)\to N(q) \phi(\Delta 
p_z)$ and can be calculated using the usual Lorentz invariant Feynman diagram techniques.

\subsection{CP violation in production}
So far we have been looking at $\chi\to N$ transition. However if there are more than one species of $\chi, N$ then the couplings $Y$ become in general 
complex matrices, and, if it contains a physical phase, this can lead to CP violating processes. The Lagrangian \eqref{eq:Toy_model_1} generalises to 
\bea 
\label{Eq:example-mod}
\mathcal{L} =
i \bar \chi_i P_R  \sl\d \chi_i + i \bar N_I \sl\d N_I- M_I \bar N_I N_I- Y_{iI} \phi \bar N_I P_R\chi_i  - y_{I\alpha}(h\bar l_\alpha ) P_R N_I + h.c.
\eea
where $h$ and $l_\alpha$ are the usual SM Higgs and fermions that we couple to the heavy $N_I$, $P_R, P_L$ are the  chiral projectors. 
We choose this assignment of chirality in agreement with our further toy models.
In
particular the rates $\Gamma(\chi_i\to 
N_I)\neq\Gamma(\bar \chi_i\to \bar N_I) $ 
and after the phase transition there 
could be an asymmetry in $N,\bar N$ and
$\chi, \bar \chi$ populations. Let us calculate these asymmetries. It is known that at tree level no asymmetries can be generated since both processes will be proportional to $|Y_{iI}|^2$, so we need to consider one loop corrections to it, in particular it is known that the imaginary part of the loop is the crucial ingredient for asymmetry generation. In general performing such calculation in the presence of the bubble wall background is quite involved, however things simplify if the 
bubble expansion is ultra-relativistic. Then we can expand in $\frac{\vev{\phi}}{E}\sim \frac{\vev{\phi}}{\gamma T}$ parameter. In this case it will be sufficient, similarly to the tree level result, to focus only on the effects in the matrix element at ${\cal O}\l(\frac{\vev{\phi}}{\gamma T}\r)$ i.e. one scalar VEV insertion.

\subsubsection{Calculation of the light $\to$ heavy transition \emph{at 1-loop level}}
\label{sec:asymmetry}
Let us now compute the asymmetries in the populations of the various particle immediately after the 
PT in the case of the model in Eq.\eqref{Eq:example-mod}. First of all we need to know the CP violating effects in the $\chi_i\to N_I$
transition, which will appear in the interference of the loop and tree level diagrams. 
\bea
&&A(\chi_i\to N_I)_{\rm tree}\propto Y_{iI}\nn
&&A(\chi_i\to N_I)_{\rm 1-loop}\propto\sum_{k,J} Y_{iJ}Y_{k J}^*Y_{k I}\times f^{(\chi \phi)}_{IJ}+\sum_{\alpha,J} Y_{iJ}y_{\alpha J}^*y_{\alpha I}\times f^{(h l)}_{IJ}
\eea
where the functions $f^{(hl)}$ and $f^{(\chi\phi)}$ refer to the loop diagrams with virtual $\chi,\phi$ and $hl$ respectively. As a consequence, there will be the following asymmetries in $N_I$ populations immediately after the PT
\bea
 && \epsilon_{Ii} \equiv  \frac{| \mathcal{M}_{i \to I}|^2 -| \mathcal{M}_{\bar i\to \bar I}|^2}{\sum_i | \mathcal{M}_{i \to I}|^2 +| \mathcal{M}_{\bar i \to \bar I}|^2} \nn\\
 &&=\frac{2\sum_{k,J} {\rm Im} (Y_{iI} Y_{iJ}^*Y_{k J}Y^*_{k I} ) {\rm Im }f^{(\chi\phi)}_{IJ}}{ \sum_{i}|Y_{i I}|^2}+\frac{2\sum_{\alpha,J} {\rm Im} (Y_{iI} Y_{iJ}^*y_{\alpha J}y^*_{\alpha I} ) {\rm Im}f^{(h l)}_{IJ}}{ \sum_{i}|Y_{i I}|^2},
 \label{eq:CP_asym_1}
\eea
where $\epsilon_{Ii}$ refers to asymmetry in $N_I$ particle population which are produced from the $i$ initial flavour of $\chi_i$. The loop functions take the form 
\bea 
f^{(hl)}_{IJ}(x) \equiv 2\int \frac{d^4p}{(2\pi)^4}\frac{P_R\slashed{p}P_L(\slashed{p}_{out}+M_J)P_L}{(p^2+i\epsilon)((p-p_{out})^2+i\epsilon)(p_{out}^2-M_I^2+i \epsilon)} 
\\
f^{(\chi \phi)}_{IJ}(x) \equiv \int \frac{d^4p}{(2\pi)^4}\frac{P_L\slashed{p}P_R(\slashed{p}_{out}+M_J)P_L}{(p^2+i\epsilon)((p-p_{out})^2+i\epsilon)(p_{out}^2-M_I^2+i \epsilon)} ,
\label{eq:loop-functions}
\eea
{where $p,p_{out}$ are the initial (particle $i$) and final (particle $I$) state four momenta.
The factor of two in front of the $f^{(hl)}_{IJ}(x)$ function comes from the two contributions with a loop of $\nu_L, h^0 $ and $e_L,h^+$. This factor is absent in the case of $f^{(\chi \phi)}_{IJ}(x)$ because we have only the loop of $\chi , \phi$. 
The imaginary part of those loop functions take the form
\begin{align}
\label{Loop-functions}
& \text{Im}[f^{(hl)}_{IJ}(x)] =\frac{1}{16\pi} \frac{\sqrt{x}}{1-x}, \qquad x = \frac{M_J^2}{M_I^2}
\\
& \text{Im}[f^{(\chi \phi)}_{IJ}(x)] = \frac{1}{32\pi} \frac{1}{1-x}.
\end{align}

Summing over the
flavours of $\chi_i$ we arrive at the following asymmetry in $N_I$ abundance\footnote{The asymmetry can be equivalently obtained from the ``tree-level" graph of the 1PI effective action by integrating out the fermions. }
\bea
\epsilon_I\equiv \sum_i \epsilon_{Ii}=
\frac{2\sum_{\alpha,J,i} {\rm Im} (Y_{iI} Y_{iJ}^*y_{\alpha J}y^*_{\alpha I} ) {\rm Im}f^{(h l)}_{IJ}}{ \sum_{i}|Y_{i I}|^2}.
\label{eq:CP_asym_2}
\eea

Note that the only diagrams contributing to the asymmetry are shown on the Fig.\ref{fig:diaglepto}) and these have  virtual $hl$. 
\begin{figure}
    \centering
    \includegraphics{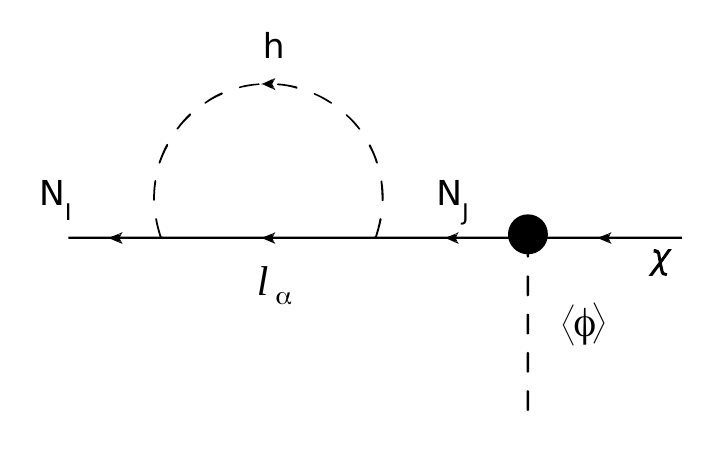}
    \caption{The diagram contributing to the function $f^{(hl)}$}
    \label{fig:diaglepto}
\end{figure}

So far we have shown that  during the production we can  create a difference in the abundances of $N_I$ and $\bar N_I$ inside the bubble. However since it was produced by $1\to 1$ transitions exactly the same difference will be present inside the bubble also for the abundances of $\bar \chi_i$ and $\chi_i$. Using the ``$\vev{\phi}\neq 0$" and ``$\vev{\phi}=0$" subscribes for the particle densities inside and outside the bubbles and taking into account that the number density of some particle entering inside the bubble is $n = \frac{\Delta N}{\Delta A \Delta t} \frac{\Delta t}{\Delta z} = \frac{J}{v_w}$ with entering flux $J = \int \frac{p_z d^3p}{p_0(2\pi)^3} f_\chi (p, T_{\text{nuc}}) $ we can conclude that, in the plasma frame,
\bea
&&n_{N_I}^{\vev{\phi}=0}(T_{\text{nuc}}) \simeq  0, \hbox{~~~~Boltzmann suppressed}
\eea
and
\bea
\label{eq:n_prod}
n_{N_I}^{\vev{\phi}\neq 0}&\simeq &
 \frac{1}{\gamma_w v_w}  \int \frac{d^3p}{(2\pi)^3} P_{\chi \to N}(p) \times f_\chi (p, T_{\text{nuc}}) 
\nn
& \simeq &
 \sum_i\frac{|Y_{iI}|^2 \vev{\phi}^2}{ M_{I}^2 \gamma_w v_w}  \int \frac{d^3p}{(2\pi)^3}  \times f^{eq}_\chi (p,T_\text{nuc})\Theta ( p_z- M_{I}^2/\vev{\phi})\nn
 &\simeq&
\sum_i \frac{ |Y_{iI}|^2 }{\pi^2 \gamma_w^3 v_w}\times  \frac{\vev{\phi}^2T_\text{nuc}^2}{M_{I}^2}\bigg( \frac{M_{I}^2/\vev{\phi}}{1-v_w}+ \frac{T_\text{nuc}(2-v_w)}{\gamma_w (v_w-1)^2}\bigg) \times e^{- \gamma_w \frac{M_{I}^2}{\vev{\phi}} \frac{1-v_w}{T_\text{nuc}}}\nn 
&
=& \sum_i\frac{ |Y_{iI}|^2 T_\text{nuc}^3 \vev{\phi}^2}{\pi^2M_I^2}  e^{-  \frac{M_I^2}{2\vev{\phi}T_\text{nuc} \gamma_w }}  + \mathcal{O}(1/\gamma_w) 
 \nn
 &\simeq & \sum _i\theta_{iI} ^2 n_{\chi^i}^{\vev{\phi}=0}(T_{\text{nuc}}).
\eea
where $v_w = \sqrt{1-1/\gamma_w^2} \approx 1 - \frac{1}{2\gamma_w^2}$ is the velocity of the wall.
The integral is performed in the wall frame and the $\gamma_w^{-1}$ factor in front  takes care of the conversion to the plasma frame.
In the second line we introduced the expression Eq.\eqref{eq;trans_light_heavy} of the probability of light to heavy transition and assumed that, the transition being a detonation, the density distribution of $\chi$ is the equilibrium distribution $f^{eq}_\chi \approx e^{-\frac{{\gamma_w}(E_\chi - v_wp^\chi_z)}{T_\text{nuc}}}$ (using Boltzmann distribution as a simplifying assumption) and $E_\chi = \sqrt{p_z^2 + \vec{p}^2_\perp}$. In the third line we performed the phase space integral.
In the last approximation, we have taken the exponential to be unity since the wall is relativistic and $\gamma_w T_{\rm nuc} \gg M_I$, and defined 
\bea 
\theta_{iI}\equiv \frac{|Y_{iI}| \vev{\phi}}{M_I}.
\eea
This means that some of the abundance of $\chi_i$ has been removed from the plasma and since we are focusing on $1\to 1$ transitions, this gives : 
\bea
&&\sum_I \Delta n_{N^I} = - \sum_i \Delta n_{\chi^i} \Rightarrow \nn
&&\sum_I \l(\Delta n_{N^I}-\Delta n_{\bar N^I} \r)= - \sum_i \l(\Delta n_{\chi^i}-\Delta n_{\bar\chi^i}\r) ,
\eea
where $\Delta n_{N,\chi}$ are the differences in abundances of the particles in the broken and unbroken phases.
As a consequence, there will be also an asymmetry in the abundances of the light fields. Note that the asymmetry in the $\chi$ field will be further diluted by the factor $\sim \frac{y^2 \vev{\phi}^2}{M^2}$ due to the large symmetric thermal densities of the  light fields during and after the phase transition. 
\begin{table}[]
\begin{center}
\begin{tabular}{llllll}
\hline
\multicolumn{1}{|l|}{} & \multicolumn{1}{l|}{$N_I$} & \multicolumn{1}{l|}{$N_I^c$} & \multicolumn{1}{l|}{$\chi_i$} & \multicolumn{1}{l|}{$\chi_i^c$}   & \multicolumn{1}{l|}{$\Delta n_{N^I}$}   \\ \hline
\multicolumn{6}{|l|}{ ~~~~~~~~~~~~~~~~~~~~~~~~$\epsilon=0$, without CP-violation}                                                                                               
\\ \hline
\multicolumn{1}{|l|}{Out} & \multicolumn{1}{l|}{0} & \multicolumn{1}{l|}{0} & \multicolumn{1}{l|}{$n_{\chi^i}$} & \multicolumn{1}{l|}{$n_{\chi^i}$} & \multicolumn{1}{l|}{0} 
\\ \hline
\multicolumn{1}{|l|}{In} & \multicolumn{1}{l|}{$\theta^2_{Ii}n_{\chi^i}$} & \multicolumn{1}{l|}{$\theta^2_{Ii}n_{\chi^i}$} & \multicolumn{1}{l|}{$(1-\theta^2_{Ii})n_{\chi^i}$} & \multicolumn{1}{l|}{$(1-\theta^2_{Ii})n_{\chi^i}$} 
& \multicolumn{1}{l|}{0} 
\\ \hline
\multicolumn{6}{|l|}{ ~~~~~~~~~~~~~~~~~~~~~~~~$\epsilon\neq 0$, with CP-violation}                                                                                               
\\ \hline
\multicolumn{1}{|l|}{Out} & \multicolumn{1}{l|}{0} & \multicolumn{1}{l|}{0} & \multicolumn{1}{l|}{$n_{\chi^i}$} & \multicolumn{1}{l|}{$n_{\chi^i}$} & \multicolumn{1}{l|}{0} \\ \hline
\multicolumn{1}{|l|}{In} & \multicolumn{1}{l|}{$\theta^2_{Ii}(1- \epsilon_{Ii})n_{\chi^i}$} & \multicolumn{1}{l|}{$\theta^2_{Ii}(1+ \epsilon_{Ii})n_{\chi^i}$} & \multicolumn{1}{l|}{$(1-\theta^2_{Ii}(1- \epsilon_{Ii}))n_{\chi^i}$} & \multicolumn{1}{l|}{$(1-\theta^2_{Ii}(1+ \epsilon_{Ii}))n_{\chi^i}$}& \multicolumn{1}{l|}{$2\epsilon_{Ii} \theta^2_{Ii} n_{\chi^i}$} \\ \hline
\label{ref-values}
\end{tabular}
\end{center}
\caption{Densities and asymmetry, with and without CP-violation, inside and outside of the bubble. For clarity we got rid of the temperature dependence, assuming that the density have to be evaluated at the nucleation temperature. } 
\end{table}

 \section{Application of the mechanism for baryogenesis}
 \label{sec:applications}
 In the previous section we have shown that the wall, if it becomes relativistic enough, can produce states much heavier than the reheating temperature and also that this production process can induce CP-violation via the interference of tree-level and loop-level diagrams. 
 
Now we will present some examples of applications of this new CP-violating source for the explanation of the observed matter asymmetry. Of course, many other examples could take advantage of the configuration presented in the previous section, so what we will present now serve as a proof of existence.\footnote{It is also clear that the baryogenesis model can be built from CP-violating decay of the produced heavy particle due to the bubble expansion. This is nothing but the non-thermal baryogenesis~\cite{Lazarides:1991wu, Asaka:1999yd, Hamaguchi:2001gw,Barbier:2004ez, Dimopoulos:1987rk, Babu:2006xc,McKeen:2015cuz, Aitken:2017wie, Elor:2018twp,  Grojean:2018fus, Hamada:2018epb, Pierce:2019ozl,Asaka:2019ocw}. In the models in this paper, this component of the asymmetry production is negligible.
} For this reason, in the following we will present two classes of models that take advantage of the mechanism present above.

 \subsection{Phase-transition induced leptogenesis}
 \label{sec:model1}
\begin{figure}
 \centering
 \includegraphics[scale=0.25]{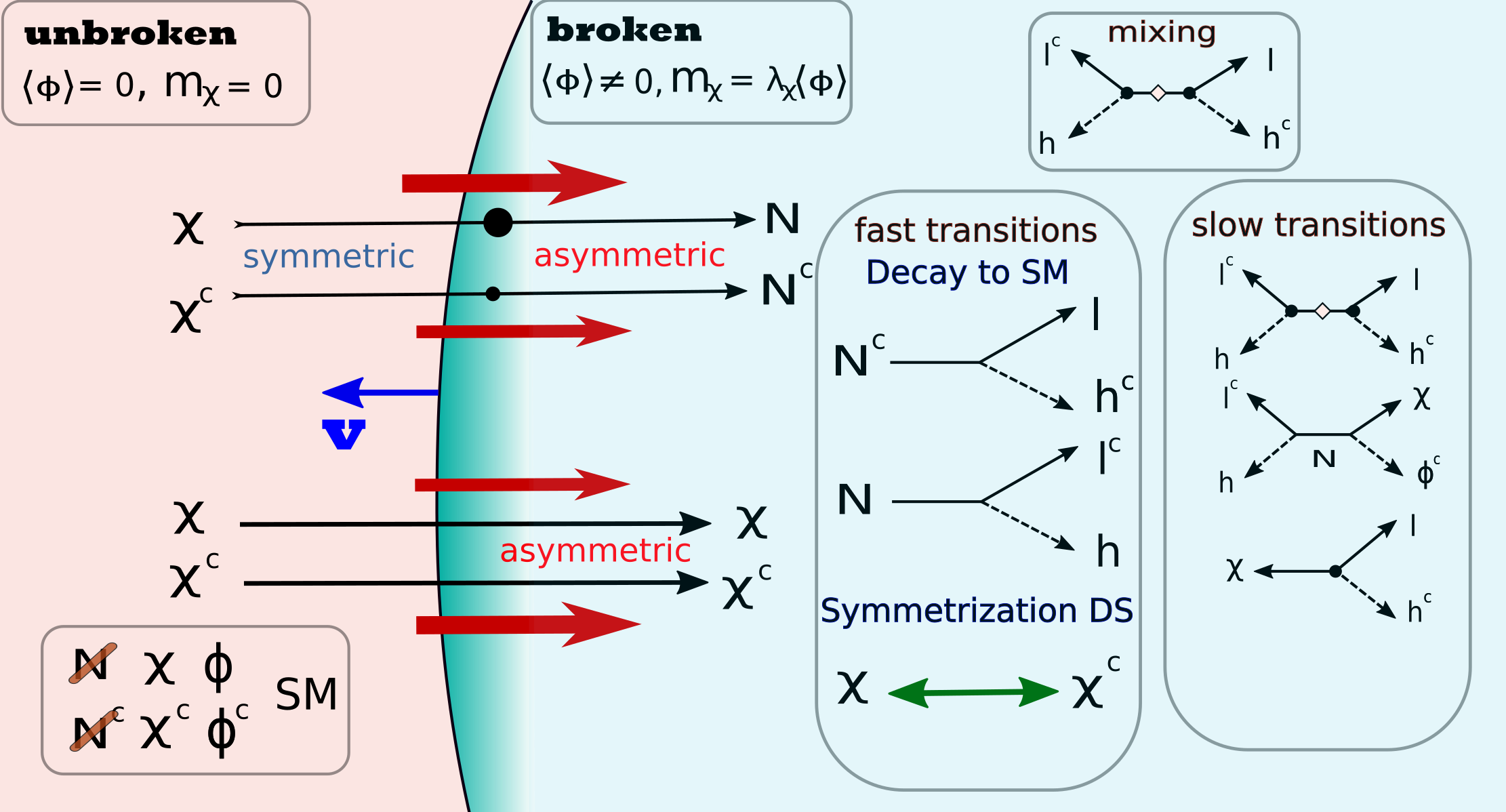}
 \caption{Mechanism at play in the phase transition-induced leptogenesis. In the diagrams the black dot denotes a mixing between $\chi$ and $N$ and so the insertion of a factor of $\theta$, the white diamond is a $\chi$ mixing insertion. A thicker arrow designates a larger flux (though it is exaggerated on the figure.) }
 \label{fig:diagram_1}
 \end{figure}
Let us consider the following extension of the Lagrangian in Eq.\ref{Eq:example-mod}, where we have introduced $\phi$ -dependent Majorana mass for the field $\chi$ and kept the rest of the interactions the same. 
We restrict to only one specie of the Majorana fermion $\chi$, since it is sufficient for the generation of CP phase.
\bea
\label{Eq:ToyMod}
\mathcal{L}_{\text{int}}& = & \underbrace{\sum_{I} \bigg(Y_{I}(\phi^\dagger \bar{\chi}) P_L N_{I} + Y_{I}^{\star}\bar{N}_{I}P_R(\phi \chi)   \bigg) - V(\phi) + \frac{1}{2}\lambda_\chi \phi \bar{\chi}^c \chi + \sum_{I} M_I \bar N_I N_I}_{\text{Toy model of Dark Sector}} 
\\ \nonumber
&+ &\underbrace{\sum_{\alpha I} y_{\alpha I} (h \bar{l}_{\alpha,SM})P_R N_{ I} + h.c.}_{\text{Connection to SM}},
\eea 
The interactions in Eq. \ref{Eq:ToyMod} respect $U(1)$ lepton number  with the following  charge assignments $L(\chi) = -1, L(N) = 1$ and $L(\phi)=2$.  This symmetry is obviously broken after the phase transition by the  VEV of $\vev{\phi}$ field and the Majorana mass of the $\chi$ field  $m_\chi = \lambda_\chi \vev{\phi}$.
Perturbativity bound imposes that $\lambda_\chi\lesssim \sqrt{4\pi}$. This model is only an example of the realisation of our scenario, we list some alternatives in Appendix \ref{sec:variations}.
The generation of the baryon asymmetry proceeds as follows: during the bubble expansion we generate asymmetry in $N$ and $\chi$, as they have been estimated in the section \ref{sec:asymmetry}. Immediately after the transition, the asymmetry in $\chi$ is washed out due to the lepton-number violating 
Majorana mass term, which constitutes the first source of asymmetry for the system. Part of this asymmetry in $N$ is passed to the SM lepton sector during the decay $N\to l h $, which constitutes a second source of asymmetry for the system, via the usual CP-violating decay. We will see that the dominant contribution depends on the different couplings of the systems. This asymmetry in return is passed to the 
baryons by sphalerons, similarly to the original leptogenesis models \cite{Fukugita:1986hr}. The scheme of the construction is shown on the Fig.\ref{fig:diagram_1}.

\subsubsection{Estimating the baryon asymmetry}
So far we have generated the asymmetry in $N$ particle population, however we need to find  what part of it will be passed to SM lepton sector.
This can be done by comparing the branching ratios of $N\to hl$ and $N\to \chi \phi$ decays
 \bea
 \label{eq:asym_model_1}
 \frac{n_{l}-n_{l^c}}{s} &\simeq & \frac{1}{s(T_{reh})}
\sum_{iI}\epsilon_{Ii} \frac{3\zeta(3) |Y_{iI}|^2 T_\text{nuc}^3 \vev{\phi}^2}{4\pi^2M_I^2} \times 
\frac{Br(N_I \to hl)}{Br(N_I \to hl)+Br(N_I \to \chi \phi)}
\nn
 &\simeq &
 \frac{ 135 \zeta(3) g_\chi}{8 \pi^4 g_\star}\sum_I \theta^2_I
 \frac{2\sum_{\alpha,J} {\rm Im} (Y_{I} Y_{J}^*y_{\alpha J}y^*_{\alpha I} ) {\rm Im}f^{(h l)}_{IJ}}{ |Y_{ I}|^2} \bigg(\frac{T_{nuc}}{T_{reh}}\bigg)^3\times 
{\frac{\sum_\alpha | y_{\alpha I}|^2}{\sum_\alpha |y_{\alpha I}|^2+{ | Y_{I}|^2}}}\nn
 \eea
where $g_*$ is total number of degrees of freedom and $s(T) = \frac{2\pi^2}{45}g_\star T^3$, and $g_\chi$ is the number of degrees of freedom of $\chi$ particle. The reheating temperature, $T_{reh}$, is the temperature of the plasma immediately after the end of the transition, when the latent heat of the transition warmed up the plasma. If the transition is instantaneous, we can estimate $\frac{\pi^2 g_{\star}[T_{\rm reh}]}{30}T_{\rm reh}^4= V[0]-V[v]+\frac{\pi^2 g_{\star}[T_{\rm nuc}]}{30}T_{\rm nuc}^4$ where $V$ should be understood as the thermally corrected potential at the transition. By assuming $O(1)$ parameters in the potential and dominant latent heat, $T_{\rm reh}\sim v.$  
The factor $\theta_I^2$ is the suppression due to the heavy field production and $\l(\frac{T_{\rm nuc}}{T_{reh}}\r)^3$ factor come from the fact that $n_N$ is fixed by the nucleation temperature (see Eq.\eqref{eq:n_prod}) and $s$ is at the reheating temperature after the PT. The factor ${\frac{\sum_\alpha | y_{\alpha I}|^2}{\sum_\alpha |y_{\alpha I}|^2+{| Y_{I}|^2}}}$ appears since a part of the asymmetry in $N$ is decaying back to $\phi \chi$, thus washing out a part of the asymmetry.
\paragraph{Lepton asymmetry generation in decay} Note that there is an additional effect contributing to the baryon asymmetry generation. The decays of the heavy fields $N$ are out of equilibrium and there is 
a CP phase in the Yukawa interactions. Thus the rates
\bea
\Gamma(N\to \bar h l)\neq \Gamma (\bar N\to  h \bar l )
\eea
induce a non-vanishing CP-violation in decay
\bea
\label{eq:decay-asym}
\epsilon^I_{decay}\equiv \frac{\Gamma(N^I\to \bar h l)-\Gamma (\bar N^I\to  h \bar l )}{\Gamma(N^I\to \bar h l)+\Gamma (\bar N^I\to  h \bar l )}.
\eea
As a consequence, after the asymmetry induced by the production of heavy states, there will be an additional asymmetry due to the decay scaling as : 
\bea
\l.\frac{n_{l}-n_{l^c}}{s}\r|_{decay}\sim 
\sum_I\frac{\theta^2_I}{g_\star} \epsilon_{decay}^I \bigg(\frac{T_{nuc}}{T_{reh}}\bigg)^3\times 
{\frac{\sum_\alpha | y_{\alpha I}|^2}{\sum_\alpha |y_{\alpha I}|^2+{ | Y_{I}|^2}}}
\eea
where the asymmetry in decay $\epsilon^I$ will be generated by the diagram similar to the one in Fig.\ref{fig:diaglepto} with $h,l$ in the final state and $\phi, \chi$ inside the loop. In the limit $m_\chi,m_\phi\ll M_I$ (which is exactly where one VEV insertion approximation used in the section \ref{sec:CPviol} is motivated) the loop function for both production and decay will be exactly the same up to the factor of 2 (particles running in the loop are EW singlets) compared to Eq.\eqref{eq:CP_asym_2}.
However the couplings will be complex conjugates so that
\bea
\epsilon^I_{decay}=
-\frac{{\rm  Im}[Y_{I}Y^*_{J} y_{\alpha_J}y^*_{\alpha_I}]{\rm Im}[f^{(hl)}_{IJ}]}{\sum_{\alpha}|y_{\alpha I}|^2}.
\eea
Combining both effects and taking into account sphaleron rates converting the lepton asymmetry to the baryon, we obtain the following baryon asymmetry:
\bea
\frac{\Delta n_B}{s} \equiv \frac{n_{B}-n_{\bar B}}{s} \simeq&& -\frac{28}{79}\times \frac{135 \zeta(3) g_\chi}{8\pi^4 g_*}\times \sum_I{\theta^2_I} \sum_{\alpha,J} {\rm Im} (Y_{I} Y_{J}^*y_{\alpha J}y^*_{\alpha I} ) {\rm Im}f^{(h l)}_{IJ}\nn
&&\times \l( \frac{2}{|Y_{ I}|^2}-\frac{1}{\sum_{\alpha} |y_{\alpha I}|^2}\r)\bigg(\frac{T_{nuc}}{T_{reh}}\bigg)^3
{\frac{\sum_\alpha | y_{\alpha I}|^2}{\sum_\alpha |y_{\alpha I}|^2+{ | Y_{I}|^2}}}
\label{eq:asym_model_1}
 \eea
The prefactor $-\frac{28}{79}$  comes 
from the sphalerons rates (see  \cite{Harvey:1990qw}). The observed asymmetry is given by $\frac{\Delta n_B}{s} \sim 8.8\times 10^{-11} $. The quantities in Eq.\eqref{eq:asym_model_1} can be 
estimated in the following way; 
outside of the resonance regime but 
for mild hierarchy between the 
masses of the heavy neutrinos $M_1\lesssim M_2\lesssim M_3$, $2{\rm Im}f^{(h l)}_{IJ} \to \frac{1}{8\pi}$, $g_\star \sim 100$ and ${ \sum_\alpha |y_{\alpha I}|^2 \gtrsim  |Y_{ I}|^2}$ induces ${\frac{\sum_\alpha | y_{\alpha I}|^2}{\sum_\alpha |y_{\alpha I}|^2+{|Y_{I}|^2}}} \sim 1$. As a consequence, the production of the observed asymmetry demands, in order of magnitude,
\bea
\text{Max}[\theta^2 y^2] \bigg(\frac{T_{nuc}}{T_{ reh}}\bigg)^3 \sim 10^{-6} .
\eea
with an $O(1)$ CP phase and $|Y_I| \sim |Y_J|$.

\color{black}
\subsubsection{Constraints on the model}

Let us examine various bounds on the construction proposed. Let us start from neutrino masses. Indeed
after the PT, the Lagrangian \eqref{Eq:ToyMod} generates a dimension 5 operator of the see-saw form 
\cite{Minkowski:1977sc,Yanagida:1979as,GellMann:1980vs,Glashow:1979nm,Mohapatra:1979ia}
\bea 
\sum_{I ,\alpha, \beta}\theta_{I}^2\frac{y_{\alpha I}y^*_{\beta I}(\bar{l}^c_\alpha h)(l_\beta h )}{m_\chi}
\eea
which induces a mass for the {active} neutrinos (for the heaviest light neutrino)
\bea
\text{Max}[m_{\nu}] \sim \text{Max}\bigg[\sum_I|y_{\alpha I}|^2\theta^2_{I}\bigg] \frac{  v_{EW}^2 }{m_\chi}.
\label{eq:neutrino_mass}
\eea
Combining Eqs. \eqref{eq:asym_model_1}, 
\eqref{eq:neutrino_mass} with 
observed neutrino mass scale and the constraints
$\text{Max}[\theta^2_{I}] \gtrsim 10^{-5}, y \sim \mathcal{O}(1)$, we obtain the following constraints
\bea 
\Rightarrow \qquad  m_\chi \gtrsim 5\times 10^{9} \text{GeV}  \qquad \Rightarrow \qquad  \vev{\phi} \gtrsim 10^{9} \text{GeV}.
\label{constraints_1}
\eea
Let us list additional conditions on this baryogenesis scenario which must be satisfied. First of all, 
the decay processes of $\chi \to h l$, and  $\chi \to (h  l)^*$ have the same probability since  $\chi$ is a Majorana fermion.     
Then we need to make sure that, immediately after the reheating, processes involving $\chi$ do not erase the asymmetry stored in the SM sector. Let us list these processes and their rates: 
 \begin{itemize}
 \item $ \chi$  production in $lh$  collisions:
Ideally we have to solve the Boltzmann equation for the density evolution, which focusing only on this process will be given by:
\bea
s z H(z)\frac{dY_{l, l^c}}{dz}=-\frac{Y_{l,l^c}Y_h}{Y^{eq}_{l,l^c}Y^{eq}_h}\gamma(h l \to\chi)+\frac{Y_{\chi}}{Y_\chi^{eq}}\gamma(\chi\to l h, (l^c h)),
\eea
where $z\equiv m_\chi/T$ (not to be confused with the spatial direction z along the wall) and $Y_i\equiv n_i/s$.
However note that $\chi,\chi^c$ 
decay  quickly with the rate 
$\Gamma\sim \frac{y^2 \theta^2 
m_\chi}{4\pi}\gg H$ unless we consider 
the scales close to the Planck mass, 
this process induces that the 
density $Y_\chi$ is always kept
close to equilibrium. Introducing the 
asymmetry density 
$Y_{\Delta_\alpha}\equiv 
Y_{l_\alpha}-Y_{l_\alpha^c}$ and 
subtracting for the matter anti-matter densities, we obtain
\bea
s z H(z)\frac{d Y_{\Delta_\alpha}}{dz}=-\frac{Y_{\Delta_\alpha}}{Y_{l_\alpha}}\gamma(hl_{\alpha}\to \chi)
\eea
where $\gamma_\alpha[z] \equiv \gamma(hl_{\alpha}\to \chi)$ is given by\cite{Davidson:2008bu}
\bea
\gamma_\alpha[z]=\frac{g_\chi T^3}{2\pi^2}z^2 K_1(z)\Gamma_\alpha
\eea
where the Bessel functions $K_1(z)$ satisfy the two limiting behaviours
\bea 
zK_1(z) = \begin{cases}
1 \qquad z\ll 1,
\\
\sqrt{\frac{\pi z}{2}}e^{-z}\qquad z\gg 1.
\end{cases}
\label{eq_limit_cases}
\eea
So,  for large values of $z$, we get
\bea
\label{eq:condition2to1}
\frac{d Y_{\Delta_\alpha}}{d z}\simeq -\frac{0.42 e^{-z}z^{5/2}}{g_*^{1/2}g_\alpha}\l(\frac{M_p}{m_\chi}\r)\l(\frac{g_\chi \Gamma_\alpha}{m_\chi}\r)Y_{\Delta_\alpha}, \qquad \Gamma_{\alpha} \approx \l|\sum_{I} y_{\alpha I}\theta_{I} 
\r|^2\frac{m_\chi}{8\pi g_\chi}.
\eea
Solving this equation numerically we can find that $Y_{\Delta_\alpha}$ remains invariant for $m_\chi/T_{reh}\gtrsim 15$ (for the scale $m_\chi\sim 10^{9}$ GeV), so that the wash out process can be safely ignored. The following approximate relation for the minimal $m_\chi/T_{reh}$ to avoid wash out is valid
\bea 
\frac{m_\chi}{T_{reh}}  \gtrsim \log \frac{M_p}{m_\chi} 
-9
\label{eq:hierarchy}
\eea
where we took $\theta_I \sim 10^{-2}$ as a typical value.
Similarly to the process above there will be additional effects which can lead to the wash-out of the lepton asymmetry like; $hl \to \phi \chi$.
 However the rate of this reaction will be further suppressed by the phase space and it will be subleading compared to the $hl\to \chi$.

The mild hierarchy in Eq.\eqref{eq:hierarchy} between $m_\chi$ and the reheating temperature is easily achievable in the case of long and flat potentials where the difference of energies between false and true vacua is smaller than the VEV: $T_{reh}\sim (\Delta V/g_*)^{1/4} \red{\lesssim} {\cal O}(10^{-1}) \vev{\phi}\sim  {\cal O}(10^{-1}) m_\chi$, which can be achieved for example by simply taking small quartic coupling in the $\phi$ potential. This happens typically in models with approximate conformal symmetry
\cite{Creminelli:2001th,Nardini:2007me,Konstandin:2011dr, Azatov:2020nbe}, and in the case of models containing heavy fermions\cite{Carena:2004ha,Angelescu:2018dkk}.
\item On top of these wash out effects there will be the ``usual" processes from the $llhh$ operator $h^c l\to hl^c$, $ll \to  hh $ and $hh \to l l$  which will violate the lepton number with rates 
\begin{align}
\Gamma(h^c l_\alpha \to hl_\beta^c )(T)& = \frac{2g_\beta}{\pi^4}\sum_{ i I }\frac{\theta_{iI}^4}{m_\chi^2}y^2_{i\alpha}y^2_{i\beta} \frac{n_h}{n^{eq}_h}\frac{1}{n^{eq}_\alpha}T^6 \approx \frac{4}{1.2 \pi^2g_\alpha }\sum_{ i I }\frac{\theta_{iI}^4}{m_\chi^2}y^2_{i\alpha}y^2_{i\beta} T^3.
\label{eq:hlhl}\nn
\Rightarrow\Gamma(h^c l\to hl^c )&\approx \frac{2}{1.2\pi^2 }\l(\frac{m_\nu}{v_{EW}^2}\r)^2T^3.
\end{align}
(where we consider the heaviest light neutrino $m_\nu$ in our estimates)
and may wash out the asymmetry created. Requiring these processes to be slow, we arrive at the condition

\bea
\Gamma(h^c l\to hl^c)< H(T_{reh}) \quad \Rightarrow \quad
T_{reh}\lesssim 5\sqrt{g_\star}\frac{v_{EW}^4}{M_p m_\nu^2}\sim 5\times 10^{12
} \text{ GeV}.
\eea
where we took $m_\nu^2 \sim 0.0025 \text{ eV}^2$. 
\item During the symmetry breaking topological defects may be formed. For the cosmic strings $\vev{\phi}\lesssim 10^{14}$ GeV is needed to evade the CMB bound~\cite{Charnock:2016nzm}. If the $U(1)$ is explicitly broken by the potential of $\phi$, domain walls will form. Depending on the explicit breaking the domain wall or string network would be unstable and decay. {In this case the CMB bound is absent. Instead, the string-wall network emits gravitational waves and may be tested in the future with VEV $\gtrsim 10^{14}\,$ GeV and the axion mass range of $10^{-28}-10^{-18}$ eV~\cite{Hiramatsu:2013qaa,Gorghetto:2021fsn}}.

\item (Pseudo) Nambu-goldstone boson which may be identified as $\arg \phi$, exists in this scenario. If it acquires mass via the explicit breaking of the $U(1)$ symmetry, the late-time coherent oscillation should not over-close the Universe. This requirement sets an upper bound on the explicit breaking-term or the decay should happen early enough.
In the former case, we have a prediction on dark radiation corresponding to the effective neutrino number of $\Delta N_{\rm eff}\sim 0.03$, since the light boson is easily thermalized  around and after the PT. This can be tested in the future. 
\end{itemize}

In conclusion we can see that this 
construction can lead to the viable 
baryogenesis if there is a mild 
hierarchy between the scales; $M_I > \vev{\phi}$ and $m_\chi,M_I> T_{\rm reh}$. In particular we need $M_I/\vev{\phi}\gtrsim 10$ in order to remain in the range of validity for our calculation from perturbation 
theory point of view and we need $(m_\chi,M_I)/T_{reh}\gtrsim 15$ to 
suppress the wash-out. Correct 
reproduction of neutrino masses makes 
this mechanism  operative in the range of scales $10^9 < \vev{\phi} < 5\times 10^{12}$ GeV. We would like to emphasize that the discussion above assumed one mass 
scale for all $M_I$, and similarly all of the couplings $Y, y$ are of the same scale. However this is not the case in general and the discussion of such ``flavour" effects can significantly modify the allowed scale of the transition.

Before going to the next model let us mention that  there is no lepton number  violation in the symmetric phase, and in  the  broken  phase  $\chi$  is  heavier  than  the  plasma temperature. Thus the thermal leptogenesis does not happen in the parameter range for this scenario.

\subsection{Low-energy baryogenesis via EW phase transition}
\label{sec:model2}
\begin{figure}
 \centering
 \includegraphics[scale=0.25]{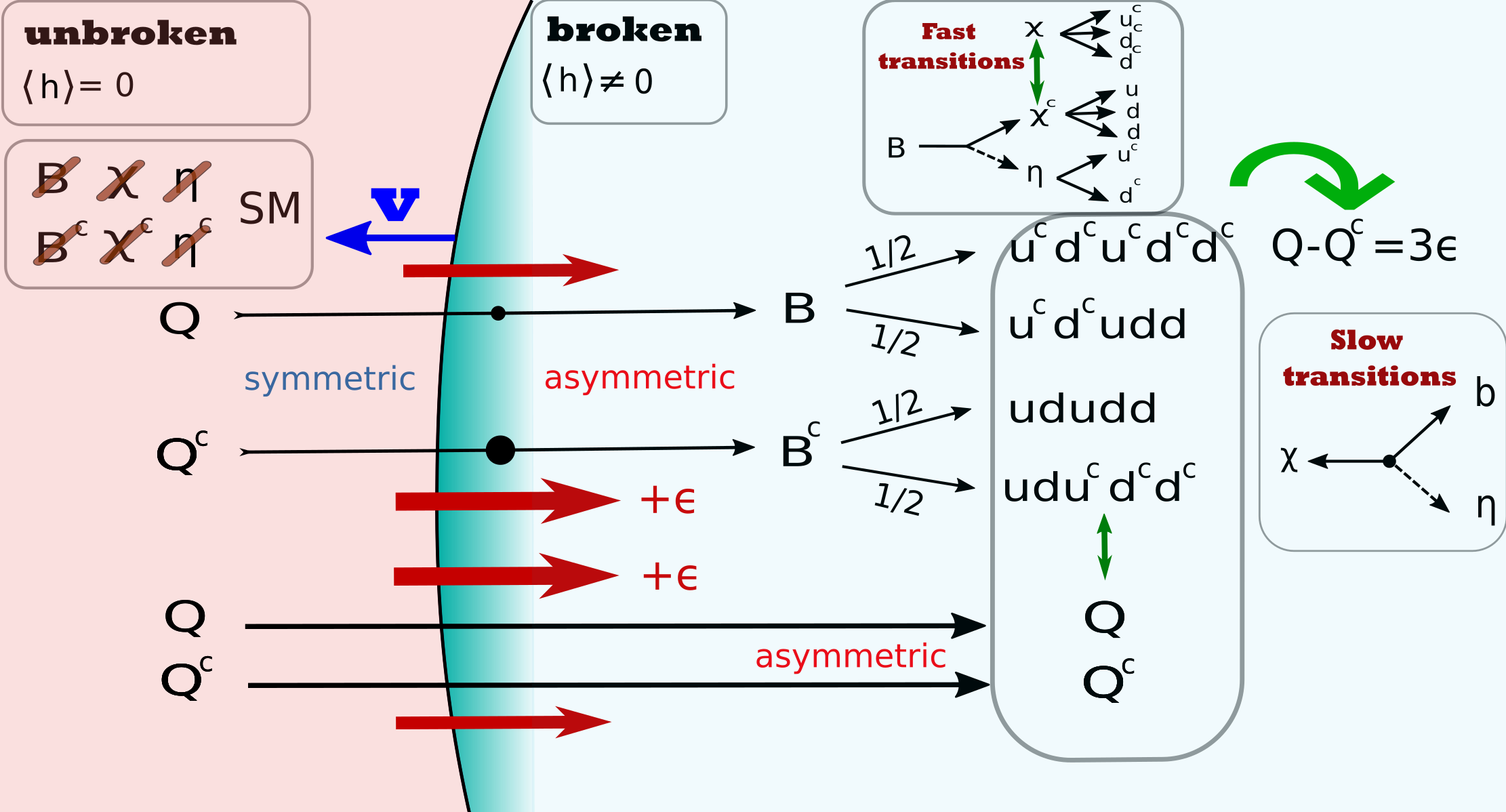}
 \caption{Mechanism at play in the low energy baryogenesis.  }
 \label{fig:diagram_2}
 \end{figure}
In the previous section we have presented a model  generating the baryon asymmetry during the phase transition at the high scale. However we can wonder whether the mechanism proposed (in section \ref{sec:model1}) can be effective for generation of the baryon asymmetry during the EW phase transition.  The necessary ingredient  for the mechanism is a strong first order electroweak phase transition and various studies 
indicate that even a singlet scalar
(see ref.\cite{PhysRevD.45.2685,Choi:1993cv,Espinosa:1993bs,Profumo:2007wc,Espinosa:2011ax,Chen:2017qcz,Ellis:2018mja}) or dimension six operator (see ref. \cite{Chala:2018ari,Huang:2015izx,Bodeker:2004ws,Delaunay:2007wb,Ellis:2018mja})
extensions of SM can do the job. In this paper however we take an agnostic approach about the origin of such EW FOPT and just assume that it has happened with nucleation and reheating temperature as an input parameters
and leave the  detailed analysis of the explicit realizations for the future studies. Below we present a prototype model:
\bea
\label{eq:modelB}
{\cal L} =&&{\cal L}_{SM} +  m_\eta^2 |\eta |^2+\sum_{I=1,2} M_{I} \bar B_I B_I \nn
+&&\l(\sum_{I=1,2}Y_I (\bar B_I H)P_L Q+
{y_I \eta^*\bar B_I P_R\chi}
+ \kappa \eta^c d u
 +\frac{1}{2} m_\chi \bar{\chi^c} \chi + h.c. \r) .
\eea
As before we do not write kinetic terms. 
The model contains a Majorana field $\chi$ and two  vector-like $B$ quarks with the masses $M_{1,2}\sim m_\chi$.
Here $\eta$ is a scalar field which is in the fundamental representation of QCD with electric charge $Q(\eta)=1/3$, $Q,u,d$  are  the  SM quark doublet and singlets respectively, we ignore the flavour indices for now, 
$H$ is the SM Higgs  and we assume that the EW phase transition is of the first order with relativistic enough bubbles\footnote{
This model is not the only possible realisation of the successful baryogenesis via the EW phase transition. In Appendix \ref{sec:variations}, we list several variations of this model.}.
Note that the interaction  $ L H \chi^c$ is consistent with all the gauge symmetries of the model, however we set it to zero in order to avoid proton decay. This can be attributed to some accidental discrete symmetry. 

Let us assume that only the third generations couples to the heavy vector like $B$ quark, $Q=(t,b)$ in Eq.\eqref{eq:modelB}, then unlike the previous leptogenesis model, 
asymmetry will be generated when the relativistic SM $b$ quarks are hitting the wall.

Let us look at the baryon number assignments of the various fields in our lagrangian:
$B(\eta)=2/3, B(\chi)=1$, so that the $m_\chi$ violates the baryon symmetry by two units. 
In this case, the story goes as follows: the sweeping of the relativistic wall, via the collision of the b-quarks with bubbles, produces $B_I, B_I^c$. Thus inside the bubble
\bea
n_{B_I}-n_{B^c_{I}}=-\theta_{I}^2 \epsilon_I n_b^0\nn
n_{b}-n_{b^c}=\sum_I\theta_{I}^2 \epsilon_I n_b^0
\label{eq:asymmB}
\eea
where $n_b$ is the number density of the bottom-type quark, $\theta_I \approx \frac{Y_I \vev{H}}{M_I}$ is the mixing angle  and  $\epsilon_I$ is defined like in  Eq.
\eqref{eq:CP_asym_1} (in this case there is no $i$ index since we coupled it only to  the third generation of quarks). CP asymmetry will be generated by the diagram represented on the Fig.\,\ref{fig:loop_2}
with $\chi,\eta$ fields running inside the loop. 
\begin{figure}
\centering
\includegraphics{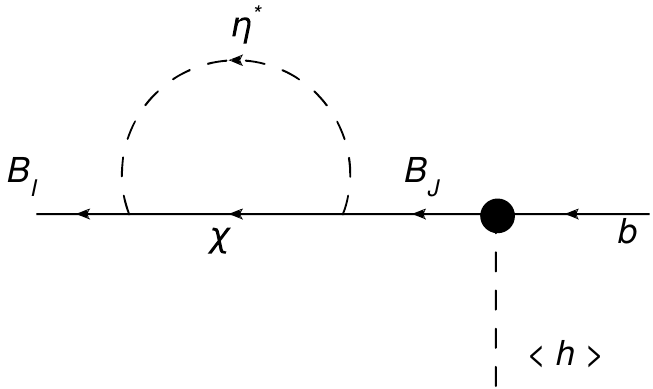}
\caption{One loop diagram contributing to the $b\to B$ transition }
    \label{fig:loop_2}
\end{figure}
The loop function generated by the diagram of Fig. \ref{fig:loop_2} becomes
\bea 
{f_B^{IJ}(x)} =
 \int \frac{d^4p}{(2\pi)^4}\frac{P_R(\slashed{p}+m_\chi)P_L(\slashed{p}_{out}+M_J)P_L}{(p^2-m_\chi^2+i\epsilon)((p-p_{out})^2-m_\phi^2+i\epsilon)(p_{out}^2-M_I^2+i \epsilon)} 
\eea
after taking the imaginary part, we obtain
\bea 
{\text{Im}[f^{IJ}_B(x)]} =
\frac{1}{32\pi}
\frac{M_I M_J}{M_I^2-M_J^2}\frac{\sqrt{(M_I^2 - m_\eta^2 + m_\chi^2)^2 - 4m_\chi^2M_I^2}}{M_I^4} 
\l( M_I^2+m_\chi^2-m_\eta^2\r) 
\label{eq:loop_func}
\eea 
Compared to  Eq.\,\eqref{Loop-functions} we have an additional $1/2$ factor (at the massless limit of $\eta,\chi$) because $\eta,\chi$ are $SU(2)_L$ singlet now.

Similarly to our general discussion after the passage through the wall the following asymmetric abundances will be generated 
\bea
\label{eq:refBb}
\sum_I  \l(n_{B_I}-n_{ B_I^c}\r)
= -( n_{b}- n_{b^c}) .
\eea
Let us see what will happen after $B_I$ decays.
There are two decay channels that lead back to b quarks and thus can erase the asymmetry, one which is direct back-decay to SM $B_I\to b h$ ($h$ is the CP even  neutral component of the Higgs doublet) and the other through $\chi, \eta$, $B_I\to \chi \eta^c $ if kinematically allowed.
The last channel will lead to the following decay chain 
\bea
\label{netprocess}
B_I\to \chi \eta^c \to \chi d^c u^c.
\eea
Let us look at the decays of the $\chi$  field. For concreteness we will assume the following ordering of the masses 
\bea
M_I> m_\chi >M_\eta.
\eea
 Then the Majorana fermion $\chi$ is not stable and decays
 \bea
 \chi \to b \eta, 
 \eea

 where $\chi^c b \eta$ interaction is generated after EWSB due to the mixing between vectorlike quarks and SM fields once the Higgs boson develops the VEV. The $\eta$ field later decays to two quarks. However the Majorana nature of the field $\chi$ makes the decay to the CP conjugate final state open as well so that
 \bea
 &&\chi \to b \eta\to b  du\nn
&& \chi\to b^c \eta^c \to b^c d^c u^c,
 \eea
 decays are allowed
and both  final states have the same probabilities. As a result there will be two decay chains of $B$ one leading to the  generation of the baryon asymmetry and another to the wash-out 
 \bea
 &&(i)~~ \text{wash-out}:~~B_I\to \chi d^c u^c \to (b d u  d^c u^c) \qquad ~~B_I^c\to \chi^c d u \to (b^c d^c u^c  d u)\nn
 &&(ii)~~\text{asymm. generation}:~~B_I\to \chi^c d^c u^c \to (b^c d^c u^c  d^c u^c) \qquad ~~B_I^c\to \chi d u \to (b d u  d u)\nn
 \label{eq:cascade}
 \eea
 As a result the asymmetry between SM quarks and antiquarks will be given 
  \begin{align} 
(n_{q}-n_{q^c}) &=\sum_I (n_{B_I}-n_{ B_I^c}) 
 \l[\l(-\frac{5}{2}+\frac{1}{2}\r)Br(B_I\to \chi \eta^c)+Br (B_I\to b h) 
\r]+(n_b-n_{b^c})\nn,
 &=-3\sum_I (n_{B_I}-n_{ B_I^c}) Br(B_I\to \chi \eta^c),
 \end{align}
where we have used $Br(B_I\to \chi \eta^c)+Br (B_I\to b h) =1$ and Eq.\ref{eq:refBb} to derive the last relation.
 At last we have to take into account CP violating decays of the $B$ particles similarly to the discussion in Eq.\eqref{eq:decay-asym}
 \bea
 \epsilon^I_{decay}
=
  \frac{\Gamma(B^I\to \chi \eta^c)-\Gamma(B^{I,c}\to \chi^c \eta)}{\Gamma(B^{I,c}\to \chi \eta^c)+\Gamma(B^{I,c}\to \chi^c \eta)}
 =-\frac{4 {\rm Im} (Y_I Y^*_J y_I^*y_J) {\rm Im} [f_B^{IJ}]|_{m_{\eta,\chi}\to 0}}{|y_I|^2}
 \eea
 where the loop function is exactly equal to the one in Eq. \eqref{eq:loop_func} with zero masses of the particles inside the loop and an extra factor of 2, since inside the loop will now circulate the EW doublet.
 Thus for the total baryon asymmetry we obtain
 \bea
 \frac{\Delta n_{Baryon}}{s}  
\approx&&\frac{135 \zeta(3)}{8 \pi^4} \sum_{I,J}\theta_I^2   \frac{|y_I|^2}{|y_I|^2+|Y_I|^2}\times \frac{g_b}{ g_\star}\bigg(\frac{T_{\rm nuc}}{T_{reh}}\bigg)^3\nn
&&
\times {\rm Im}  (Y_I Y^*_J y_I^*y_J )
\l(-\frac{2 {\rm Im} [f_B^{IJ}]}{|Y_I|^2}+\frac{4 
{\rm Im} [f_B^{IJ}]|_{m_{\chi,\eta}\to 0}
}{|y_I|^2}
\r).
 \eea

With $\frac{135\zeta (3) }{8\pi^4} \times \frac{|y_I|^2}{|y_I|^2+|Y_I|^2}\times \frac{g_b}{ g_\star} \frac{{\rm Im} [Y_2 Y_1^* y_2^* y_1]}{16\pi |Y_I|^2} \sim 10^{-(3-4)}$ {and $g_b=6$ being the degrees of freedom of $b$ quark} and assuming $y_I=O(1)$ and  $|y_I|\gg |Y_I|\sim |Y_J|$, recovering the observed baryon abundance $\frac{\Delta n_{Baryon}}{s} \sim 8.8\times 10^{-11} $ requires, in order of magnitude,
\bea 
\Rightarrow \boxed{\theta_I^2 \bigg(\frac{T_{\rm nuc}}{T_{reh}}\bigg)^3 \sim 10^{-(6-7)}}.
\eea

 The decay chains described above are very fast compared to the Hubble scale so that we can treat them as instantaneous. Indeed the  
 \bea
 &&\Gamma(\chi\to \eta b) \approx \l|\sum_{I} y_I \theta_I\r|^2\frac{m_\chi(1-m_\eta^2/m_\chi^2)}{8\pi g_\chi} \Rightarrow\\
 &&\frac{ \Gamma(\chi\to \eta b) }{H(v_{EW})}\simeq 10^{-3}\frac{M_p}{m_\chi} \gg 1, 
 \eea
since in the range of interest $\theta \lesssim 10^{-2}\Rightarrow m_\chi < 10^{2}$ TeV.

Thus, after this first phase of very fast decays, that produces the baryon asymmetry in the quark sector, slow transition mediated by the heavy states can still wash out the asymmetry. In this part, we check for which region of the parameter space it is not the case.
The various wash-out transitions include 
 \bit
 \item $b\eta \to \chi $
 \newline
The decoupling of this transition provides the following condition, reminiscent of Eq.\eqref{eq:condition2to1}. In particular writing the Boltzmann in equation for the B-asymmetry we will get:
\bea
s z H(z)\frac{d( Y_{b}-Y_{b^c})}{dz}\simeq -\gamma_{\chi\to \eta b}\l(\frac{Y_b Y_\eta-Y_{b^c}Y_{\eta^c}}{Y_{b}^{eq}Y_{\eta}^{eq}}\r)
\eea
assuming that the asymmetries in $B$ and $\eta$ are related as follows 
\bea
Y_{b,b^c}=(1\pm\epsilon_q)Y_{eq}^b,~~~
Y_{\eta,\eta^c}\simeq 
Y_{eq}^\eta
\eea
then, we arrive at the following equation
\bea
&&s z H(z)\frac{d( \epsilon_q Y_{eq})}{dz}\simeq -\gamma_{\chi\to \eta b} \epsilon_q,~~~\gamma_{\chi\to \eta b}=\frac{g_\chi T^3}{2\pi^2}z^2 K_1(z)\Gamma(\chi \to \eta b)\nn
&&\frac{d \epsilon_q}{d z}=-
\frac{0.42 e^{-z}z^{5/2}}{g_*^{1/2}g_q}\l(\frac{M_p}{m_\chi}\r)\l(\frac{g_\chi \Gamma(\chi\to \eta b)}{m_\chi}\r)\epsilon_q
\nn
&&\Gamma(\chi\to \eta b) \approx \l|\sum_{I} y_I \theta_I\r|^2\frac{m_\chi(1-m_\eta^2/m_\chi^2)}{8\pi g_\chi} \label{washout}
\eea
Then the process is decoupled for the temperatures of EW scale if $m_\chi/T_{reh}\gtrsim 30$, which pushes us to the limits of the maximal asymmetry which we can achieve in the mechanism. Indeed assuming $T_{reh}\sim 100$ GeV, we are required to have $m_\chi \gtrsim 3$ TeV and $m_B\gtrsim 3$ TeV. 
Note that on top of the process above there will be  reactions  $\eta b\to \eta^c b^c $ which will also lead to the wash out of the asymmetry. 
This process is suppressed by the Boltzmann factor for $\eta$ field abundance so that the condition to not erase the asymmetry becomes
\bea
\frac{m_{B,\chi,\eta}}{T_{reh}}\gtrsim 30.
\eea
\item $ddu  \leftrightarrow d^c d^c u^c $
\newline
After integrating out all the new heavy fields the following baryon violating number operator  is obtained.
\bea
\frac{ddu \overline{d^c d^c u^c}}{M_\eta^4}\times \frac{1}{m_\chi}\times \theta^2
\qquad \Rightarrow \qquad
{\frac{1}{4\pi^5}\bigg(\frac{1}{16\pi^2}\bigg)^2}\frac{T_{reh}^{11} }{M_\eta^8 m_\chi^2}\theta^4\lesssim \frac{T_{reh}^2}{M_{p}}.
\eea
However, it can easily be seen that the rate of the baryon number violating processes mediated by it are much slower than the Hubble expansion as well.
\eit
\subsubsection{Experimental signatures}
This low-energy model has the interesting consequence that it induces potential low-energy signatures. In this section, we enumerate those possible signatures without assuming that $Q, u,\text{and } d$ are the third generation quarks.  
\paragraph{ Neutron oscillations}
The baryon number violating processes in the model will violate the baryon number only by $2$ units, so proton decay is not allowed but $n-\bar n$ oscillations can be present\cite{Fridell:2021gag}.
Integrating out the heavy states, we obtain the following operator
 \bea 
\frac{1}{\Lambda^5_{n\bar{n}}}\overline{u^c d^cd^c} udd  \equiv\frac{(\sum{\kappa \theta_I y_I})^2}{M_\eta^4 m_\chi }\overline{u^c d^cd^c} udd 
\eea

thus inducing a neutron mixing mass of the form
\bea 
\delta m_{\bar{n}-n} \sim \frac{\Lambda_{QCD}^6}{M_\eta^4 m_\chi}(\sum{\kappa \theta_I y_I})^2.
\eea
 Current bounds on this mixing mass are of order $ \delta m_{\bar{n}-n}  \lesssim 10^{-33} $ GeV~\cite{BaldoCeolin:1994jz, Abe:2011ky,Rao:1982gt, Buchoff:2012bm,Syritsyn:2016ijx}.
This is extremely significant if the SM quarks in the Lagrangian are in first or second generation. 
 If we take order 1 couplings and $\theta^2 \sim 10^{-5}$, it places a bound on the typical mass scale of
\bea 
\label{nnbar}
\Lambda_{n \bar n}\gtrsim 10^{6} {\rm GeV}~~~(M_\eta,m_\chi) \gtrsim 10^5 \text{GeV} 
\eea
This bound becomes weaker if the new particles couple only to the third  generation. Then we have an additional suppression factor $(V_{td}^4V_{bu}^2)^{1/5} \sim 10^{-2}$.
As a consequence, depending on the flavor of $Q,u,d$ our scenario can be tested in the future experiment~\cite{Phillips:2014fgb,Milstead:2015toa, Frost:2016qzt, Hewes:2017xtr}. 

\paragraph{ Flavor violation}
The model predicts new particles in the $1-10^3$ 
TeV range coupled to the SM light quarks- $\eta$ 
field. So the question about low energy bounds 
naturally rises. However the FCNC are absent at 
tree level for the $\eta$-diquark field   \cite{Giudice:2011ak}. The loop level effects can lead to strong constraints if $\eta d u$ coupling contains the light generation quarks \cite{Giudice:2011ak}, but if $d u$ are only the third generation fields $t_R, b_R$ the bounds are practically absent.

\paragraph{ Bounds from EDMs}
 If there is CP violation in the mixing between bottom quark and heavy bottom partners then we expect an operator of the form {
 \bea
 -i \frac{g_3\tilde d_q}{2} \bar Q \sigma^{\mu \nu}T^A\gamma_5 QG^A_{\mu \nu}
 \eea
 which is the chromo-electric dipole moment (see \cite{Engel:2013lsa} for a review)
 with  coefficient scaling like
 \bea 
\tilde d_q \sim \text{Im}[ y_I^2] \frac{\theta_I^2 m_b}{16\pi^2} \frac{1}{\Lambda_{EDM}^2}
 \eea
 where $\Lambda_{EDM} \sim M_\eta \sim m_\chi \sim (1-100)$ TeV is the scale of the new physics that we are considering.}
  Up-to-date bounds \cite{Chang:1990jv,Gisbert:2019ftm} are $\tilde d_b < 1.2 \times 10^{-20} cm\sim 10^{-6} GeV^{-1}$ , 
 {which include the nucleon EDM bound via the RG effect.}
 Taking typical values of the mixing angle $\theta^2 \sim 10^{-4}$, it can be seen that those bounds are not stringent.

 Electron EDM is known to be one of the important test of the EW baryogenesis theories. In our model the leading contribution appears at three loop level due to the Barr-Zee type \cite{Barr:1990vd} of diagram with $b-B$ mixing. The estimate of the dipole operator goes like 
 \bea 
 \frac{d_e}{e} \sim \frac{m_e (y Y e)^2}{(4\pi)^6}\bigg(\frac{1}{\Lambda_{EDM}^2}\bigg) \sim 3\times 10^{-33}\times \l(\frac{10 {\rm TeV}}{\Lambda_{EDM}}\r)^2 {\rm cm}
 \eea 
 which is four orders of magnitude below the current experimental bound
 \cite{Andreev:2018ayy} $|d_e|< 1.1\times 10^{-29} {\rm cm} \cdot e$ .

}

\paragraph{ Gravitational waves}

One very robust prediction of such a scenario is the large amount of GW emitted at the transition, with peak frequency fixed by the scale of the transition $f_{\text{peak}}\sim 10^{-3}\frac{T_{\text{reh}}}{ \text{GeV}}$ mHz (see \cite{Weir:2017wfa} for review). Such SGWB signal could be detected in future GW detectors such as LISA\cite{amaroseoane2017laser,Caprini:2019egz}, eLISA\cite{Caprini:2015zlo}, LIGO\cite{vonHarling:2019gme, Brdar:2018num}, BBO\cite{Corbin:2005ny, Crowder:2005nr}, DECIGO\cite{Seto:2001qf, Yagi:2011wg, Isoyama:2018rjb}, ET\cite{Hild:2010id,Sathyaprakash:2012jk,Maggiore:2019uih}, AION\cite{Badurina:2019hst}, AEDGE\cite{Bertoldi:2019tck}. This array of observers will be able to probe GW with frequencies in the window of mHz to kHz, which is the optimal scale for this mechanism to take place.

\paragraph{ Direct production in colliders}
An almost coupling- and flavor-independent bound is the LHC one. The heavy quark can be produced via the strong or electro-magnetic interaction. From the recent squark or gluino bounds in the LHC e.g.\cite{Aad:2020aze, Sirunyan:2019ctn}, we expect a mass bound of $\sim 2$ TeV on the lightest colored particle.  
 
\subsubsection{Parameter region} 
\begin{figure}
 \centering
 \includegraphics[scale=0.45]{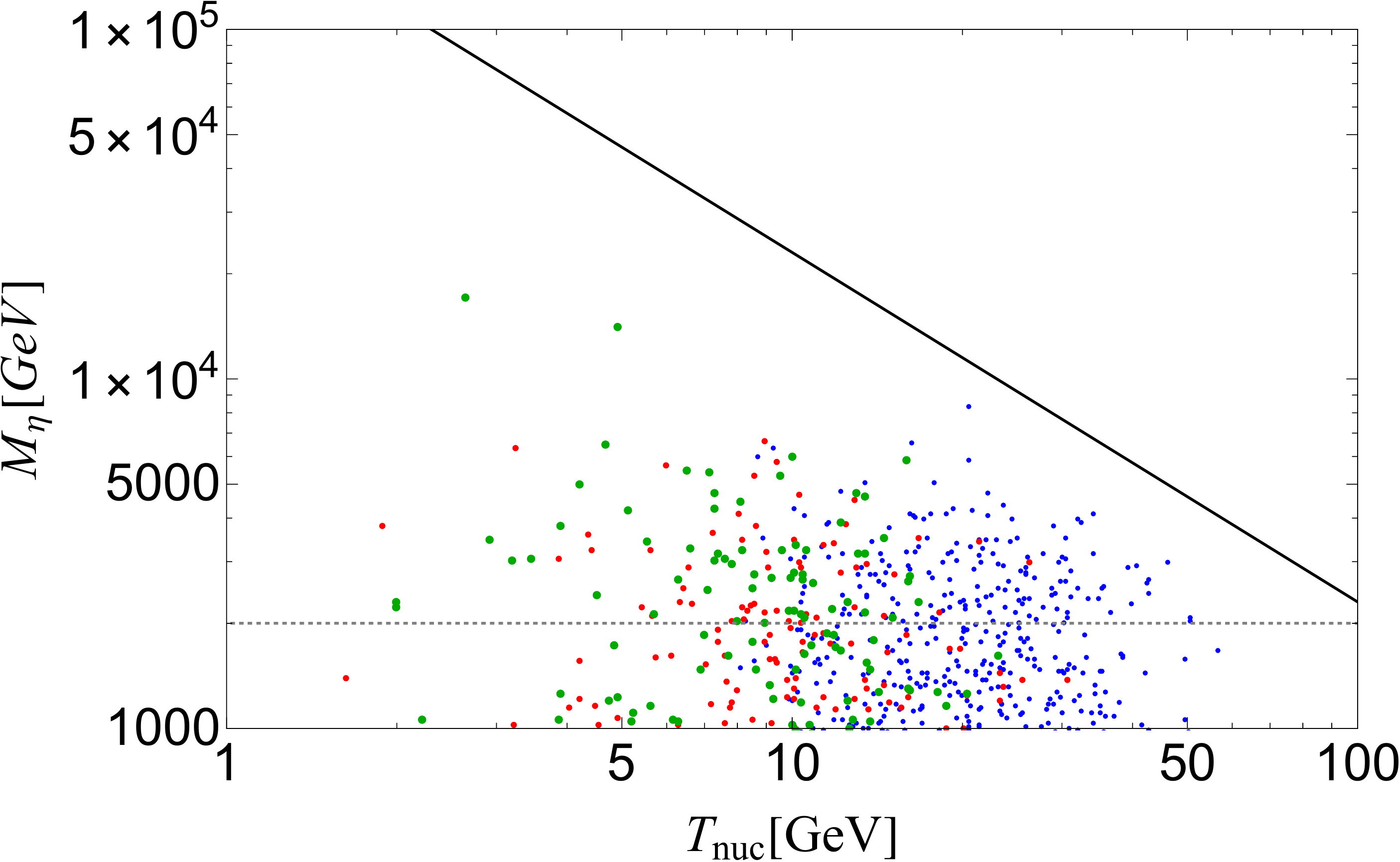}
 \caption{ $M_{\eta}$ (which is the lightest colored particle mass) vs $T_{\rm nuc}$ in the low energy baryogenesis. 
The green points satisfy \eqref{nnbar},
  while the red and blue points do not satisfy it and thus require a special flavor-structure. Both (the lighter of) $M_1$ and $M_2$ are taken to 
  satisfy the conditions from maximal wall velocity \eqref{Mmax}, which is shown by the black solid line, in the appendix for green and red (blue) points. 
 Here we fix $\vev{H}=T_{\rm reh}=100\,$GeV to consider the electroweak phase transition and fix $\Delta n_B/s =8.8 \times 10^{-11}$. Other parameters are randomly chosen within the perturbative unitarity range.  Below the 2TeV (dotted line) may be disfavored by the null detection of new colored particle in the LHC.  } 
 \label{fig:res}
 \end{figure}

 By combining all the previous bounds and by numerically solving the washout condition~Eq.\eqref{washout} we show the parameter region of this scenario in Fig.\ref{fig:res}. 
 Here the horizontal axis is the $T_{\rm nuc}$ which, in order to produce a relativistic bubble wall, is favored to be smaller than $T_{\rm reh}$, which we took equal to $100$ GeV. The vertical axis denotes $M_{\eta}(< M_{I}),$ which is the lightest diquark in this scenario.
 All the data points give the correct baryon asymmetry $\Delta n_B/s=8.8 \times 10^{-11},$ with all the couplings smaller than $\sqrt{2\pi}.$ The green points satisfy 
 the bound on $n-\bar n$ oscillation in Eq.\eqref{nnbar}. 
 The red and blue points do not satisfy it, which implies a special flavor structure for example that only $b$  is coupled to the BSM particles. For both red and green points, both $M_1$ and $M_2$ satisfy the relativistic wall condition \eqref{Mmax} in the appendix, which is denoted by the black solid line. For the blue points, the lighter of $M_{1}$ and $M_2$ satisfies the condition. 
The horizontal dotted at 2 TeV is the typical bound on new colored particles.
 Therefore our mechanism predicts light quark, which may be searched for in the LHC and future colliders. 
 Moreover, since our data points include parameter space both consistent and inconsistent with 
 the 
 neutron-anti-neutron oscillation bound, some points with BSM particles that also couple to the first two generation quarks can be tested in the future. 
Note again that $T_{\rm nuc}\sim O(0.1)T_{\rm reh}$ may be the consistent range for our scenario to work.

\subsubsection{Baryogenesis from non-EW FOPT}
At last we note that, with a simple modification of the model in Eq.\eqref{eq:modelB}, the mechanism can be operative for an arbitrary phase transition. Indeed let us assume that $\phi$ is the field experiencing the FOPT then the following lagrangian : 
\bea
{\cal L}_{\phi} ={\cal L}_{SM} + \sum_{I=1,2}\tilde Y_I (\bar B_I \phi) b_R+M_{I} \bar B_I B_I + \lambda_I B_I \chi^c \eta + \kappa \eta^c d u
 + \frac{m_\chi}{2} \bar{\chi}^c \chi
\eea
can induce the required baryon asymmetry.
The phenomenology remains similar to the model of Eq.\eqref{eq:modelB} however the experimental constraints from $n-\bar n$ oscillations and other low energy searches become even weaker. At the limit, the only robust experimental signature of such a scenario is the GW background emitted if the VEV is not extremely larger than the EW scale.\footnote{
For the values the VEV $\vev{\phi}\gtrsim 10^{12}$ GeV,  there is no need to suppress the interaction $H L \chi$  since the bounds from proton decay become compatible with experiment.
}

\section{Summary}
\label{sec:summary}
In this paper we have presented a novel 
mechanism for the generation of the baryon 
asymmetry during the early universe evolution. We have first shown that the mechanism of production
of the heavy particles from the 
relativistic bubble expansion during the
FOPT can lead to CP violating effects. This mechanism of particle production is
out of equilibrium, so that if baryon number violating interactions are present a baryon asymmetry can be generated.
We have constructed two viable baryogenesis models implementing this idea.  The first scenario  is 
the phase-transition-induced leptogenesis, where  
the bubble wall should be composed by some new 
Higgs field charged under the lepton number, and  
after the phase transition we are still in the 
symmetric phase of the EW interactions but in the 
broken phase of the lepton number. Later EW spharelons transfer the lepton asymmetry to the baryon sector. 
In this scenario, a net $B-L$ asymmetry is generated since the Majorana term of the right-handed neutrino after the PT violates the $B-L$ symmetry. Neutrino mass observations  make the mechanism viable if the scale of the phase transition is between $[10^9,10^{12}] $ GeV, which makes it borderline detectable for the future gravitational wave experiments such as ET\cite{Hild:2010id,Sathyaprakash:2012jk,Maggiore:2019uih}.
The second scenario can happen during EW phase transition. In this case, new 
fields charged under QCD with $\lesssim 100$ TeV scale masses are needed to have enough baryon number production. 
In both cases the baryon/lepton number 
violating interactions are coming from the Majorana masses of new heavy particles. 
This leads to the Majorana  neutrino masses in the first class of models and $n-\bar n$ oscillations in the second class of models.
Another feature of this mechanism is that baryogenesis happens for the ultra-fast 
bubble expansions thus generically strong stochastic 
gravitational wave signatures are 
expected.  
Moreover, in the case of the second scenario, the frequency range is well 
within the reach of the current and future
experiments. 

{\it \underline {Note added}} : when this paper was close to completion we have become aware of another project discussing the baryon asymmetry generation due to the heavy particles production in phase transition \cite{Baldes:2021vyz}.

\section*{Acknowledgements}
AA in part was supported by the MIUR contract 2017L5W2PT. WY in part was supported by JSPS KAKENHI Grant Nos. 16H06490, 19H05810 and 20H05851. AA and MV would like to thank the organizers of the "Computations That Matter 2021" virtual workshop for the fruitful atmosphere and  many stimulating discussions. 

\appendix

\section{Dynamics of the transition}
\label{app:dynamics}
In our discussion of the phase transition we 
have been completely agnostic regarding the 
origin of the potential and treated the Lorentz 
factor $\gamma_w$ as a free input parameter. 
This is obviously not the case and the nature of
the phase transitions as well as the dynamics of the bubble expansion depend on the details of the potential and the particle content. However even without going into explicit models we can make few semi-quantitative claims on whether the values of $\gamma_w$ needed for the heavy particles production can be achieved or not. In this respect it is important to recall the forces acting on the wall, which determine the velocity of the expansion of the bubbles. First of all there is a driving force which is given by the potential differences between true and false vacua
\bea
\hbox{Driving force}=V_{false}-V_{true}\equiv\Delta V, 
\eea
and then there are friction forces due to the plasma particles colliding with the bubble walls. Generically the calculation of these effects is quite involved but for the relativistic expansions  $\gamma_w \gg 1$
calculations become much simpler. At tree level (leading order LO), the pressure from the plasma on the bubble wall
\cite{Dine:1992wr,Bodeker:2009qy} has the form: 
\bea
\Delta \mathcal{P}_{\text{LO}} \to \sum_i g_i c_i \frac{\Delta m_i^2}{24}T_{\text{nuc}}^2,
\label{eq:LOpresLO}
\eea
where $T_{\rm nuc}$ is the nucleation temperature (roughly temperature when the PT occurs) and $\Delta m^2_i$ is the change of the masses of the particle $i$ during the PT and $c_i=1 (1/2)$ for bosons(fermions).
In the presence of the mixing between light particles and heavy particles, which is exactly our scenario of production of out-of-equilibrium heavy states, there is a second LO friction contribution \emph{from the mixing} \cite{Vanvlasselaer:2020niz}. For the models under consideration, this friction takes the form
\bea
\Delta \mathcal{P}^{mixing}_{\text{LO}} \to
\frac{T^2 Y^2}{48}\vev{\phi}^2\Theta(\gamma_w T_{\rm nuc}- M^2 L_w)
\label{eq:pressure-mix}
\eea
and in this section we consider that $\vev{\phi}$ is the scale of the symmetry breaking. At last in the presence of the gauge bosons which gain mass during the phase transition
the pressure receives Next-To-Leading order (NLO) correction\cite{Bodeker:2017cim}
\bea
\Delta\mathcal{P}_{\text{NLO}} &\simeq & \sum_{i} g_i g_{gauge}^3\gamma_w T_{\text{nuc}}^3 \frac{\vev{\phi}}{16\pi^2},
\label{eq:LOpresNLO}
\eea
where $g_{gauge}$ is the gauge coupling.
Unlike the LO pressure contributions this effect is growing with $\gamma_w$ thus eventually stopping the accelerated motion of the bubbles\footnote{Interestingly for the confinement phase transition\cite{Baldes:2020kam} \emph{even the LO contribution} is found to be proportional $\propto\gamma$.}. In order to estimate the maximal value of the $\gamma_w$ achievable during the FOPT let us consider the bubble expansion in two cases; i.e. with and without the $\gamma_w$ dependent friction (that is to say, with and without gauge bosons).
\begin{enumerate}
\item {\bf Friction is independent on $\gamma_w$} 
(no phase dependent gauge fields\footnote{There is a claim that even the gauge fields which do not get mass during the PT can provide $\gamma$ dependent friction, see Ref.\cite{Hoeche:2020rsg} for original calculation and \cite{Vanvlasselaer:2020niz} for criticism.}). In this case, if $\Delta V > \mathcal{P}_{LO}$, the bubbles will keep accelerating till the collision and the 
 $\gamma_w$ at collision can be estimated as
\cite{PhysRevD.45.3415,Ellis:2019oqb},  
\bea
&&\gamma_{w,\rm MAX} \simeq \frac{2 R_*}{3 R_0}\l(1-\frac{\mathcal{P}_{\rm LO}}{\Delta V}\r), \quad R_0 \sim 1/T_{\text{nuc}}, \quad R_*\approx \frac{(8 \pi)^{1/3}v_w}{\beta(T_{\rm nuc})},~~~\beta(T)=H T\frac{d}{dT}\l(\frac{S_3}{T}\r)
\nn
&&\Rightarrow \gamma_{w,\rm MAX} \sim \frac{M_{\text{p}}T_{{\rm nuc}} }{\vev{\phi}^2}
\label{eq:MAXmass}
\eea
where $R_\star$ is an estimate for the bubble size at collision and $R_0$ is the bubble size at nucleation and $\beta$ the inverse duration parameter of the transition. 
\item
 {\bf Friction depends on $\gamma_w$:} (gauge symmetry is broken during the PT). In this case equating the friction and the driving force gives the maximal velocity of the bubble;
\bea
\Delta V = \Delta\mathcal{P}_{\text{NLO}}(\gamma_w \equiv \gamma_{w,\text{MAX}}) \qquad \Rightarrow \gamma_{w,\text{MAX}} \approx \frac{16\pi^2}{g_{gauge}^3} \bigg(\frac{\vev{\phi}}{T_{{\rm nuc}}}\bigg)^3.
\label{eq:gamma_max}
\eea
\end{enumerate}
Combining the results for the two regimes of the bubble expansion we get
\bea
\gamma_{w}^{MAX}=\hbox{Min}\l[\frac{16\pi^2}{g_{gauge}^3} \bigg(\frac{\vev{\phi}}{T_{{\rm nuc}}}\bigg)^3,\frac{M_{\text{p}}T_{{\rm nuc}} }{\vev{\phi}^2}\r]~~\hbox{if}~~\Delta V> \Delta {\cal P}_{\rm LO}.
\eea
This results in the maximal mass of the heavy states which can be produced during the PT which is given by
\bea
\label{Mmax}
M^{MAX}\sim \text{Min}\l[\frac{4\pi}{g_{gauge}^{3/2}}\frac{\vev{\phi}^2}{T_{\rm nuc}}, \frac{M_p^{1/2} T_{\rm nuc}}{\vev{\phi}^{1/2}} \r]
\eea
If we assume $T_{\rm nuc}\sim ( 0.1-0.01 )\vev{\phi}$, then for EW phase transition the maximal mass of the state we can produce becomes
\bea
M^{MAX}_{EW}\sim (10^2-10^3) v_{EW} \sim (10-100) {\rm TeV}
\eea
So that the model in the section \ref{sec:model2} immediately satisfies this criteria. The same is true as well for the model in section \ref{sec:model1} since in this case 
\bea
M^{MAX}\sim (0.3-0.1) \sqrt{M_p \vev{\phi}}
\eea
and, for an efficient production of heavy states we needed $10^{-(2-3)}\lesssim \theta \sim \frac{\vev{\phi}}{M}$.

\section{Variations on the models}
\label{sec:variations}
In this  appendix, we will list the modifications of the models presented in the main text in Section
\ref{sec:applications}, which can realize a successful baryogenesis scenario with  significantly different phenomenology.
\subsection{Alternative Phase-transition induced leptogenesis}
The simplest modification we can take of the Eq.\eqref{Eq:ToyMod} is the consider the opposite chiralities;
\bea
\label{Eq:ToyMod2}
\mathcal{L}_{\text{int}}& = & \underbrace{\sum_{iI} \bigg(Y_{iI}(\phi \bar{\chi}_i) P_R N_{I} + Y_{iI}^{\star}\bar{N}_{I}P_L(\phi^{\dagger} \chi_i)   \bigg) - V(\phi) + \sum_i\lambda_\chi \phi \bar{\chi}^c_i \chi_i + \sum_{I} M_I \bar N_I N_I}_{\text{Toy model of Dark Sector}} 
\\ \nonumber
&+ &\underbrace{\sum_{\alpha I} y_{\alpha I} (h \bar{l}_{\alpha,SM})P_R N_{ I} + h.c.}_{\text{Connection to SM}}
\eea 
In this case the generation of CP asymmetry proceeds in the similar way to the discussion above but neutrino masses are generated at one loop level. As a result parameter space with lower masses  $m_\chi$ by around two orders of magnitude becomes accessible. We leave the thorough study of this case to further studies.

\subsection{Alternative baryogenesis models}
Let us here enumerate the possible change that we can make in the Lagrangian of Eq.\eqref{eq:modelB}
\begin{itemize}
\item We can couple $\eta$ instead of the right-handed to the left-handed ones via $\eta^c QQ$.

\item We can also couple the opposite chirality to the the diquark with the coupling $y \eta \chi P_R B$. 
In the case of the coupling $y \eta \chi P_R B$, on the top of the previous estimates, there will be an additional $m_b/M_B$ suppression in the $\chi$ decay and a similar $m_b^2/M_B^2$ is the $n-\bar n$ oscillations. 

\item We may also replace $B_i$ by an up-type-like quark, and assume that it mainly couples to the top quarks, this effect is suppressed by a further Boltzmann factor.
Then the Lagrangian is given as
\bea\sum_{i=1,2}Y_i (\bar U_i H^*) Q+M_{Bi} \bar U_i U_i + \lambda_i U_i \chi^c \eta + \kappa \eta^c dd
 +\frac{ m_\chi }{2}\bar\chi^c\chi + m_\eta  |\eta|^2.\eea Here $dd$ denotes $(d_{2} d_{3} $ or $d_{1} d_2)$. The FCNC constraint is stringent but there is a viable
parameter region. 

\end{itemize}

\bibliographystyle{JHEP}
{\footnotesize
\bibliography{biblio}}
\end{document}